\documentclass[journal,twoside,final]{IEEEtran}

\usepackage{algorithmic}
\usepackage{amsmath,amssymb,amsfonts,amsthm}
\usepackage{balance}
\usepackage{bm}
\usepackage{cite}
\usepackage{colortbl}
\usepackage[T1]{fontenc}
\usepackage{graphicx}
\usepackage{makecell}
\usepackage{mathrsfs}
\usepackage{mathtools}
\usepackage{multirow}
\usepackage{siunitx}
\usepackage{soul}               
\usepackage{textcomp}
\usepackage{threeparttable}
\usepackage[color=green!40]{todonotes}
\soulregister\cite7
\soulregister\footnote7
\soulregister\eqref7
\soulregister\ref7
\usepackage{xcolor}
\usepackage{url}
\usepackage{hyperref}
\newlength{\myvspace}
\newtheorem{lemma}{Lemma}
\newtheorem{theorem}{Theorem}

\DeclareMathOperator*{\dlb}{{\Delta}LB}
\DeclareMathOperator*{\dub}{{\Delta}UB}
\DeclareMathOperator*{\fl}{fl}
\DeclareMathOperator*{\lb}{LB}
\DeclareMathOperator*{\ub}{UB}

\newcommand{\otd}{\overline{\triangledown}}

\let\originalleft\left
\let\originalright\right
\renewcommand{\left}{\mathopen{}\mathclose\bgroup\originalleft}
\renewcommand{\right}{\aftergroup\egroup\originalright}

\graphicspath{{./}{figures/}}

\newcommand{\myauthor}{Seungyeop Kang and Kyeong Soo
  Kim,~\IEEEmembership{Senior~Member,~IEEE}}%
\newcommand{\mytitle}{Theoretical and Practical Bounds on the Initial Value of
  Skew-Compensated Clock for Clock Skew Compensation Algorithm Immune to
  Floating-Point Precision Loss}%

\begin{document}
	
\title{\LARGE \mytitle}
	
\author{%
  \myauthor%
  \thanks{%
    This work was supported in part by the Postgraduate Research Scholarships
    (under Grant PGRS1912001) and the Key Programme Special Fund (under Grant
    KSF-E-25) of Xi'an Jiaotong-Liverpool University.

    S. Kang and K. S. Kim are with the School of Advanced Technology, Xi'an
    Jiaotong-Liverpool University, Suzhou 215123, P. R. China (e-mail:
    S.Kang18@student.xjtlu.edu.cn; Kyeongsoo.Kim@xjtlu.edu.cn).%
  }%
}%

	
\maketitle

\begin{abstract}
  A clock skew compensation algorithm was recently proposed based on the
  extension of Bresenham's line drawing algorithm (Kim and Kang, \textit{IEEE
    Commun. Lett.}, vol. 26, no. 4, pp. 902--906, Apr. 2022), which takes into
  account the discrete nature of clocks in digital communication systems and
  mitigates the effect of limited floating-point precision on clock skew
  compensation. It lacks, however, a theoretical analysis of the range of the
  initial value of skew-compensated clock, which is also an initial condition
  for the proposed algorithm. In this letter, we provide practical as well as
  theoretical bounds on the initial value of skew-compensated clock based on a
  systematic analysis of the errors of floating-point operations, which replace
  the approximate bounds in Theorem~1 of the prior work.
\end{abstract}
	

\section{Introduction}
\label{sec:introduction}
\IEEEPARstart{A}{} clock skew compensation algorithm immune to floating-point
precision loss was proposed in~\cite{Kim:22-1}, which takes into account the
discrete nature of clocks in digital communication systems and thereby mitigates
the effect of limited floating-point precision on clock skew compensation based
on the extension of Bresenham's line drawing
algorithm~\cite{bresenham65:_algor}. The extended bounds on the initial value of
skew-compensated clock in Theorem~1 of \cite{Kim:22-1}, which are closely
related with the initial condition for the proposed algorithm, include a term
for floating-point operation error due to precision loss (i.e., $\varepsilon$ in
(11) of~\cite{Kim:22-1}). The value of $\varepsilon$ is not explicitly mentioned
in the theorem but approximately set for numerical examples based on the
property of the single-precision floating-point format as defined in the IEEE
standard for floating-point arithmetic (IEEE~754-2008)~\cite{IEEE:754-2008},
which may result in unexpected behaviors during the refinement of the
skew-compensated clock based on the proposed algorithm due to the lack of
guarantee of the approximate bounds on the initial condition.

In this letter, we revisit Theorem~1 of \cite{Kim:22-1} and provide practical as
well as theoretical bounds on the initial value of skew-compensated clock based
on a systematic analysis of the errors of floating-point operations.

\section{Preliminary: Relative Errors of Floating-Point Operations}
\label{sec:optimal-relative-error-bounds}
We briefly review the results of \cite{optBound} on relative errors of
floating-point operations, which are a basis for our work on both theoretical
and practical error bounds.

As the exponent range is no limiting factor in clock skew compensation, we can
define an associated set $\mathbb{F}$ of floating-point numbers with a base
$\beta$ and a precision $p$ as follows:
\begin{equation}
  \label{eq:assoc_set}
  \mathbb{F}=\{0\}\cup\left\{M\beta^{e}\big|M, e \in \mathbb{Z},
    \beta^{p-1} \leq |M| < \beta^{p}\right\}.
\end{equation}
We also denote a \textit{round-to-nearest} function by
$\fl{:}\mathbb{R}{\rightarrow}\mathbb{F}$, which satisfies
\begin{equation}
  \label{eq:round_ftn}
  |t-\fl(t)| = \underset{f\in\mathbb{F}}{\min}|t-f|, \quad t\in\mathbb{R}.
\end{equation}
Based on $\mathbb{F}$ and $\fl$, we can define two relative errors for
$t{\in}\mathbb{R}$ and $t{\neq}0$:
\begin{align}
  \label{eq:E1}
  E_{1}(t) & \triangleq \frac{\left| t-\fl(t) \right|}{\left| t \right|} \\
  \label{eq:E2}
  E_{2}(t) & \triangleq \frac{\left| t-\fl(t) \right|}{\left| \fl(t) \right|},
\end{align}
where $E_{1}$ and $E_{2}$ are the errors relative to $t$ and $\fl(t)$,
respectively; as discussed in \cite{optBound}, the relative errors may be
defined to be zero when $t{=}0$.

The optimal relative error bounds on various floating-point operations are
systematically analyzed in \cite{optBound}, and we summarize part of the results
relevant to our work in Table~\ref{tab:relative_ebs}, where $x,y{\in}\mathbb{F}$
and $u{=}\frac{1}{2}\beta^{1-p}$ is the \textit{unit roundoff} associated with
$\fl$ and $\mathbb{F}$.
\begin{table}[!tb]
  \centering
  \caption{Optimal relative error bounds~\cite{optBound}.}
  \label{tab:relative_ebs}
  \begin{tabular}{c|c|c}
    \hline
    $t$ & bound on $E_{1}(t)$ & bound on $E_{2}(t)$ \\ \hline
    real number & $\frac{u}{1+u}$ & $u$ \\
    $xy$ & $\frac{u}{1+u}$ & $u$ \\
    $x/y$ &
    $\begin{cases}
        u-2u^{2} & \text{if $\beta = 2$},\\
        \frac{u}{1+u} & \text{if $\beta > 2$}
    \end{cases}$ &
    $\begin{cases}
        \frac{u-2u^{2}}{1+u-2u^{2}} & \text{if $\beta = 2$},\\
        u & \text{if $\beta > 2$}
    \end{cases}$ \\
    \hline
  \end{tabular}
\end{table}
Note that the first row in Table~\ref{tab:relative_ebs} is for rounding a real
number $t{\in}\mathbb{R}$ to a floating-point number $x{\in}\mathbb{F}$, whose
relative errors become zero when $t{\in}\mathbb{F}({\subset}\mathbb{R})$.

\section{Clock Skew Compensation with Optimal Error Bounds}
\label{sec:clock-skew-compensation}
As in \cite{Kim:22-1}, we confine our discussions to a network with one head
node and one sensor node, where we focus only on the clock skew compensation at
the sensor node.\footnote{For instance, in the time synchronization schemes
  based on the reverse two-way message exchange, the clock offset is
  independently compensated for at the head node, while the sensor node only
  synchronizes the frequency of its logical clock to that of the reference
  clock~\cite{Kim:17-1}.} In this case, we describe the hardware clock $T$ of
the sensor node with respect to the reference clock $t$ of the head node using
the first-order affine clock model of \cite{rajan11:_joint} but without clock
offset, i.e.,
\begin{equation}
  \label{eq:hardware_clock_model_wo_offset}
  T(t) = \left(1+\epsilon\right)t,
\end{equation}
where $\epsilon{\in}\mathbb{R}$ denotes the clock skew. To compensate for the
clock skew and estimate the reference clock $t$ based on the hardware clock
$T(t)$, we need to calculate $\frac{1}{1{+}\epsilon}$ based on an estimate of
the clock skew $\epsilon$.\footnote{For detailed discussions based on a logical
  clock model, readers are referred to~\cite{Kim:22-1}.}

Because the division of floating-point numbers incurs substantial precision loss
at resource-constrained wireless sensor network (WSN) platforms with 32-bit
single precision~\cite{Kim:20-1}, we propose a clock skew compensation algorithm
using only integer addition/subtraction and comparison based on the extension of
Bresenham's algorithm in~\cite{Kim:22-1}: As discussed in \cite{Kim:22-1}, we
assume that two positive integers $D$ and $A$ are given so that $\frac{D}{A}$ be
an estimate of $\frac{1}{1{+}\epsilon}$. Theorem~1 of \cite{Kim:22-1} states
that, given a hardware clock $i$, its skew-compensated clock $j$---both $i$ and
$j$ are non-negative integers---satisfies the following condition: For
$\frac{D}{A}{<}1$,
\begin{equation}
  \label{eq:scc_bounds}
  i\frac{D}{A} - 1 - \varepsilon < j < i\frac{D}{A} + 1 + \varepsilon,
\end{equation}
where $\varepsilon~({\geq}0)$ indicates the error due to the precision loss in
computing $i\frac{D}{A}$; the value of $\varepsilon$ is approximately set to
$10^{-7}i$ for numerical examples, which, however, is not based on a systematic
and quantitative analysis of the errors of floating-point operations in
$i\frac{D}{A}$.

Lemma~\ref{lem:scc_optimal_bounds}, in this regard, provides the optimal bounds
on the initial value of skew-compensated clock based on the floating-point
operations of \textit{non-negative} real numbers.
\begin{lemma}
  \label{lem:scc_optimal_bounds}
  For $t{=}x\frac{y}{z}$, where $x,y,z{\in}\mathbb{R}$ with $x,y{\geq}0$ and
  $z{>}0$, $\fl(t)$ satisfies
  \begin{equation}
    \label{eq:scc_optimal_bounds}
    \frac{1-u+2u^{2}}{(1+u)^{2}(1+2u)}t \leq \fl(t) \leq \frac{(1+2u)^{3}(1+u-2u^{2})}{(1+u)^{2}}t.
  \end{equation}
\end{lemma}
\begin{IEEEproof}
  Because we consider rounding errors as well as multiplication and division
  errors, $\fl(t)$ is given by
  \begin{equation}
    \label{eq:scc_approx}
    \begin{split}
      \fl(t) & = \fl\left(\fl(x)\fl\left(\frac{\fl(y)}{\fl(z)}\right)\right)\\
             & =
               \left(\left(x(1+\delta_{1})\right)\left(\frac{y(1+\delta_{2})}{z(1+\delta_{3})}(1+\delta_{4})\right)\right)(1+\delta_{5})\\
             & = t\frac{(1+\delta_{1})(1+\delta_{2})(1+\delta_{4})(1+\delta_{5})}{(1+\delta_{3})},
    \end{split}
  \end{equation}
  where $\delta_{1}$, $\delta_{2}$, $\delta_{3}$ are $E_{1}$ relative errors in
  rounding a real number to a floating-point number, $\delta_{4}$ is $E_{1}$
  relative error in the division of two floating-point numbers, and $\delta_{5}$
  is $E_{1}$ relative error in the multiplication of two floating-point
  numbers. Table~\ref{tab:relative_ebs} shows that these relative errors for
  $\beta{=}2$ are bounded as follows:
  \begin{equation}
    \label{eq:rel_error_bounds}
    \begin{split}
      |\delta_{1}|, |\delta_{2}|, |\delta_{3}|, |\delta_{5}| \leq \frac{u}{1+u},\\
      |\delta_{4}| \leq u-2u^{2}.
    \end{split}
  \end{equation}

  From \eqref{eq:rel_error_bounds}, we have
  \begin{equation}
    \label{eq:error_scales}
    \begin{split}
      \frac{1}{1+u} & \leq 1+\delta_{1} \leq \frac{1+2u}{1+u},\\
      \left(\frac{1}{1+u}\right)^{2} & \leq (1+\delta_{1})(1+\delta_{2}) \leq \left(\frac{1+2u}{1+u}\right)^{2},\\
      \frac{1}{(1+u)(1+2u)} & \leq \frac{(1+\delta_{1})(1+\delta_{2})}{(1+\delta_{3})} \leq \frac{(1+2u)^{2}}{1+u},\\
      \frac{1-u+2u^{2}}{(1+u)(1+2u)} & \leq \frac{(1+\delta_{1})(1+\delta_{2})(1+\delta_{4})}{(1+\delta_{3})}\\
                    & \leq \frac{(1+2u)^{2}(1+u-2u^{2})}{1+u},\\
      \frac{1-u+2u^{2}}{(1+u)^{2}(1+2u)} & \leq \frac{(1+\delta_{1})(1+\delta_{2})(1+\delta_{4})(1+\delta_{5})}{(1+\delta_{3})}\\
                    & \leq \frac{(1+2u)^{3}(1+u-2u^{2})}{(1+u)^{2}}.\\
    \end{split}
  \end{equation}
  
  From \eqref{eq:scc_approx} and \eqref{eq:error_scales}, we obtain
  \eqref{eq:scc_optimal_bounds}.
\end{IEEEproof}

\vspace{\myvspace}%
Now we can extend Theorem~1 of~\cite{Kim:22-1} based on
Lemma~\ref{lem:scc_optimal_bounds}.
\begin{theorem}
  \label{thm:csc_with_optimal_bounds}
  Given a hardware clock $i$, we can obtain its skew-compensated clock $j$ as
  follows:
		
  \smallskip
  \noindent
  \textit{Case 1.} $\frac{D}{A}{<}1$: The skew-compensated clock $j$ satisfies
  \begin{equation}
    \label{eq:csc-scc_bounds}
    \left\lfloor{i\frac{D}{A}}\right\rfloor \leq j \leq \left\lceil{i\frac{D}{A}}\right\rceil.
  \end{equation}
  Because we cannot know the exact value of $i\frac{D}{A}$ due to limited
  floating-point precision, however, we extend \eqref{eq:csc-scc_bounds} to
  include the effect of the precision loss: For floating-point numbers with a
  base $\beta{=}2$ and a precision $p$,
  \begin{equation}
    \label{eq:csc-scc_optimal_bounds}
    \left\lfloor{\frac{1{-}u{+}2u^{2}}{(1{+}u)^{2}(1{+}2u)}t}\right\rfloor \leq j \leq
    \left\lceil{\frac{(1+2u)^{3}(1{+}u{-}2u^{2})}{(1{+}u)^{2}}t}\right\rceil,
  \end{equation}
  where $t{=}i\frac{D}{A}$ and $u{=}2^{-p}$.
  
  Let $k,{\ldots},k{+}l$ be the candidate values of $j$ satisfying
  \eqref{eq:csc-scc_optimal_bounds}. We determine $j$ by starting from the point
  $(i{-}l,k)$ and applying the extended Bresenham's algorithm with
  $\otd_{i-l}(k)$ defined in \cite{Kim:22-1} and on; $j$ is determined by the
  $y$ coordinate of the valid point whose $x$ coordinate is $i$.
		
  \smallskip
  \noindent
  \textit{Case 2.} $\frac{D}{A}{>}1$: In this case, we can decompose the
  skew-compensated clock $j$ into two components as follows:
  \begin{equation}
    \label{eq:decomposition}
    j = i\frac{D}{A} = i + i\frac{D - A}{A}.
  \end{equation}
  Now that $\frac{D{-}A}{A}{<}1$, we can apply the same procedure of Case~1 to
  the second component in \eqref{eq:decomposition} by setting $\Delta{a}$ and
  $\Delta{b}$ to $A$ and $D{-}A$, respectively. In this case, $t$ in
  \eqref{eq:scc_optimal_bounds} is equal to $\frac{D-A}{A}$. Let $\bar{j}$ be the
  result from the procedure. The skew-compensated clock $j$ is given by
  $i{+}\bar{j}$ as per \eqref{eq:decomposition}.
\end{theorem}
\begin{IEEEproof}
  \eqref{eq:csc-scc_optimal_bounds} is the result of the application of
  Lemma~\ref{lem:scc_optimal_bounds} to \eqref{eq:csc-scc_bounds}. The rest of
  Theorem~\ref{thm:csc_with_optimal_bounds} is identical to Theorem~1 of
  \cite{Kim:22-1}.
\end{IEEEproof}

\subsection{Loosening Bounds for Practical Implementation}
\label{sec:loose_bounds}
Though Theorem~\ref{thm:csc_with_optimal_bounds} provides
theoretically-guaranteed lower and upper bounds on the initial value of
skew-compensated clock $j$, obtaining the exact values of bounds---i.e., the lhs
and the rhs of \eqref{eq:csc-scc_optimal_bounds}---could be a challenge,
especially at resource-constrained sensor nodes with limited floating-point
precision.

For its practical implementation based on limited floating-point precision,
therefore, we can loosen the lhs and the rhs of
\eqref{eq:csc-scc_optimal_bounds} as follows:
\begin{align}
  \left\lfloor{\frac{1{-}u{+}2u^{2}}{(1{+}u)^{2}(1{+}2u)}t}\right\rfloor & \leq j \leq
                                                                           \left\lceil{\frac{(1+2u)^{3}(1{+}u{-}2u^{2})}{(1{+}u)^{2}}t}\right\rceil,\nonumber\\
  \left\lfloor{\frac{1{-}u}{(1{+}u)^{2}(1{+}2u)}t}\right\rfloor & \leq j \leq
                                                                  \left\lceil{\frac{(1+2u)^{3}(1{+}u)}{(1{+}u)^{2}}t}\right\rceil,\nonumber\\
  \label{eq:csc-scc_loose_bounds}
  \left\lfloor{\frac{1{-}u}{(1{+}2u)^{2}(1{+}2u)}t}\right\rfloor & \leq j \leq
                                                                   \left\lceil{\frac{(1+2u)^{3}}{1+u}t}\right\rceil,\\
  \left\lfloor{\frac{1{-}u}{(1{+}2u)^{3}}t}\right\rfloor & \leq j \leq
                                                           \left\lceil{(1+2u)^{3}t}\right\rceil.\nonumber
\end{align}

Note that the lhs and the rhs of \eqref{eq:csc-scc_loose_bounds} consist only of
the elements of $\mathbb{F}$, i.e.,
\begin{equation*}
  \begin{split}
    1 - u & = \left(2^{p} - 1\right)2^{-p} \in \mathbb{F},\\
    1 + 2u & = \left(2^{p-1} + 1\right)2^{1-p} \in \mathbb{F},
  \end{split}
\end{equation*}
which eliminates the rounding errors for those terms not belonging to
$\mathbb{F}$ in \eqref{eq:csc-scc_optimal_bounds}.

\section{Numerical Examples}
\label{sec:numerical-examples}
We first investigate the effect of floating-point precision on the calculation
of various bounds on the initial value of skew-compensated clock through
numerical experiments. Table~\ref{tab:bounds_comparison} summarizes the results
of comparison of the theoretical bounds of \eqref{eq:csc-scc_optimal_bounds}
calculated based on the floating-point formats of single precision and double precision
of IEEE~754-2008 together with the practical bounds of
\eqref{eq:csc-scc_loose_bounds} and the approximate bounds of (11)~of
\cite{Kim:22-1} based on single precision; the calculated bounds are compared to
those based on binary512 floating-point format providing 489 precision in bits,
which serve as a reference for the comparison.
\begin{table*}[!tb]
  \centering
  \begin{threeparttable}
    \caption{Comparison of Bounds on The Initial Value of Skew-Compensated
      Clock.}
    \label{tab:bounds_comparison}
    {%
      \renewcommand{\arraystretch}{1.3}
      \begin{tabular}{c|r|r|r|r|r|r|r}
        \hline
        \multirow{2}{*}{Bounds} & \multicolumn{1}{c|}{\multirow{2}{*}{$i$}} & \multicolumn{3}{c|}{$\dlb$\tnote{*}} & \multicolumn{3}{c}{$\dub$\tnote{\S}} \\
        \cline{3-8}
                                & & \multicolumn{1}{c|}{Min.} & \multicolumn{1}{c|}{Max.} & \multicolumn{1}{c|}{Avg.} & \multicolumn{1}{c|}{Min.} & \multicolumn{1}{c|}{Max.} & \multicolumn{1}{c}{Avg.} \\
        \hline
        \hline
                                & \num{1e6} & \num{0} & \num{0} & \num{0} & \num{0} & \num{0} & \num{0} \\
        Theoretical bounds of \eqref{eq:csc-scc_optimal_bounds} & \num{1e7} & \num{0} & \num{0} & \num{0} & \num{0} & \num{0} & \num{0} \\
        based on double precision & \num{1e8} & \num{0} & \num{0} & \num{0} & \num{0} & \num{0} & \num{0} \\
                                & \num{1e9} & \num{0} & \num{0} & \num{0} & \num{0} & \num{0} & \num{0} \\
        \hline
                                & \num{1e6} & \num{0} & \num{0} & \num{0} & \num{0} & \num{0} & \num{0} \\
        Theoretical bounds of \eqref{eq:csc-scc_optimal_bounds} & \num{1e7} & \num{-1} & \num{0} & \num{-5.0115e-01} & \num{0} & \num{1} & \num{2.2325e-01} \\
        based on single precision & \num{1e8} & -18 & \num{0} & \num{-7.9666e+00} & \num{0} & \num{6} & \num{1.7206e+00} \\
                                & \num{1e9} & -125 & \num{0} & \num{-4.7102e+01} & 0 & 104 & \num{3.6533e+01} \\
        \hline
                                & \num{1e6} & \num{0} & \num{0} & \num{0} & \num{0} & \num{0} & \num{0} \\
        Practical bounds of \eqref{eq:csc-scc_loose_bounds} & \num{1e7} & \num{0} & \num{2} & \num{1.0023e+00} & \num{0} & \num{1} & \num{2.2325e-01} \\
        based on single precision & \num{1e8} & \num{0} & 12 & \num{5.9980e+00} & \num{0} & 6 & \num{2.9912e+00} \\
                                & \num{1e9} & \num{0} & 120 & \num{5.9890e+01} & \num{0} & 60 & \num{2.9880e+01} \\
        \hline
                                & \num{1e6} & \num{0} & \num{1} & \num{4.9696e-01} & \num{0} & \num{0} & \num{0} \\
        Approximate bounds of (11) of \cite{Kim:22-1} & \num{1e7} & \num{-2} & \num{1} & \num{-5.0346e-01} & \num{-2} & \num{1} & \num{-2.8513e-01} \\
        based on single precision\tnote{\dag} & \num{1e8} & -20 & 10 & \num{-5.0063e+00} & -20 & 10 & \num{-4.8369e+00} \\
                                & \num{1e9} & -199 & 100 & \num{-4.9669e+01} & -199 & 100 & \num{-4.9645e+01} \\
        \hline
      \end{tabular}
    }%
    \begin{tablenotes}
    \item[*] $\dlb{=}\lb_{\text{binary512}}{-}\lb$, where
      $\lb_{\text{binary512}}$ is the lower bound of
      \eqref{eq:csc-scc_optimal_bounds} based on binary512.
    \item[\S] $\dub{=}\ub{-}\ub_{\text{binary512}}$, where
      $\ub_{\text{binary512}}$ is the upper bound of
      \eqref{eq:csc-scc_optimal_bounds} based on binary512.
    \item[\dag] With $\varepsilon{=}10^{-7}i$.
    \end{tablenotes}
  \end{threeparttable}
\end{table*}
As in \cite{Kim:22-1}, we set the value of $\varepsilon$ and $D$ to $10^{-7}i$
and $1,000,000$, respectively, and generate one million samples of $A$
corresponding to clock skew uniformly distributed in the range of
$[{-}100\,\text{ppm},100\,\text{ppm}]$.

Table~\ref{tab:bounds_comparison} shows that the double precision is enough for
the calculation of the theoretical bounds of \eqref{eq:csc-scc_optimal_bounds},
while the results for both theoretical bounds of
\eqref{eq:csc-scc_optimal_bounds} and the approximate bounds of (11) of
\cite{Kim:22-1} based on single precision violate the reference bounds of
\eqref{eq:csc-scc_optimal_bounds} based on binary512 due to the floating-point
precision loss as indicated by the negative values of $\dlb$ and $\dub$; this
implies that the results of the clock skew compensation based on Theorem~1 of
\cite{Kim:22-1} with the approximate bounds cannot be theoretically guaranteed
to be correct. Though being loose as indicated by the positive values of $\dlb$
and $\dub$, the practical bounds of \eqref{eq:csc-scc_loose_bounds} based on
single precision, on the other hand, do not violate the reference bounds in
spite of the limited floating-point precision; it turns out that the loosening
of the bounds discussed in Section~\ref{sec:loose_bounds} counteracts the effect
of limited floating-point precision on the calculation of
\eqref{eq:csc-scc_loose_bounds}.

We also compare the results of clock skew compensation algorithms again under
the same condition as~\cite{Kim:22-1}, which are summarized in
Table~\ref{tab:csc_results}.
\begin{table*}[!tb]
  \centering
  \begin{threeparttable}
    \caption{Comparison of clock skew compensation algorithms.}
    \label{tab:csc_results}
    {%
      \renewcommand{\arraystretch}{1.3}
      \begin{tabular}{c|r|r|r|r|r|r|r}
        \hline
        \multirow{2}{*}{Algorithm} & \multicolumn{1}{c|}{\multirow{2}{*}{$i$}} & \multicolumn{3}{c|}{Compensation error\tnote{*}} & \multicolumn{3}{c}{$\#$ of iterations} \\
        \cline{3-8}
                                   & & \multicolumn{1}{c|}{Min.} & \multicolumn{1}{c|}{Max.} & \multicolumn{1}{c|}{Avg.} & \multicolumn{1}{c|}{Min.} & \multicolumn{1}{c|}{Max.} & \multicolumn{1}{c}{Avg.} \\
        \hline
        \multirow{4}{*}{Single precision\tnote{\dag}} 
                                   & \num{1e6} & \num{0} & \num{0} & \num{0} & -- & -- & -- \\
                                   & \num{1e7} & \num{0} & \num{0} & \num{0} & -- & --& -- \\
                                   & \num{1e8} & \num{-4} & \num{1} & \num{-1.9842} & -- & -- & -- \\
                                   & \num{1e9} & -19 & 44 & \num{1.2502e+01} & -- & -- & -- \\
        \hline
                                   & \num{1e6} & \num{-1} & \num{0} & \num{-4.9663e-01} & \num{1} & \num{2} & \num{1.5034} \\
        Clock skew compensation with & \num{1e7} & \num{-1} & \num{0} & \num{-4.9668e-01} & \num{1} & \num{8} & \num{4.5232e+00} \\
        practical bounds of~\eqref{eq:csc-scc_loose_bounds} & \num{1e8} & \num{-1} & \num{0} & \num{-5.0220e-01} & \num{1} & 72 & \num{3.6710e01} \\
                                   & \num{1e9} & \num{-1} & \num{0} & \num{-4.9734e-01} & \num{1} & 832 & \num{4.1872e02} \\
        \hline
                                   & \num{1e6} & \num{-1} & \num{0} & \num{-4.9663e-01} & \num{2} & \num{2} & 2 \\
        Clock skew compensation with & \num{1e7} & \num{-1} & \num{0} & \num{-4.9668e-01} & \num{3} & \num{4} & \num{3.0050e+00} \\
        approximate bounds of (11) of~\cite{Kim:22-1} & \num{1e8} & \num{-1} & \num{0} & \num{-5.0220e-01} & 21 & 22 & \num{2.1005e01} \\
                                   & \num{1e9} & \num{-1} & \num{0} & \num{-4.9734e-01} & 201 & 202 & \num{2.0101e02} \\
        \hline
      \end{tabular}
    }%
    \begin{tablenotes}
    \item[*] With respect to ${\lfloor}i\frac{D}{A}{\rfloor}$ based on double
      precision.
    \item[\dag] ${\lfloor}i\frac{D}{A}{\rfloor}$ based on single precision.
    \end{tablenotes}
  \end{threeparttable}
\end{table*}
For the randomly-generated one million samples of $A$, the clock skew
compensation algorithm of \cite{Kim:22-1} with both approximate and practical
bounds provides exactly the same bounded compensation errors compared to the
unbounded errors for the single-precision algorithm, while the numbers of
iterations with the practical bounds are larger than those with the approximate
bounds for $i{\geq}\num{1e7}$. This indicates that the approximate bounds of
(11) of~\cite{Kim:22-1} could provide valid initial conditions for the cases
considered in these examples with less numbers of iterations than those for the
practical bounds of \eqref{eq:csc-scc_loose_bounds} but at the expense of the
lack of theoretical guarantee.

\section{Concluding Remarks}
\label{sec:concluding-remarks}
In this letter, we have revisited Theorem~1 of \cite{Kim:22-1} and derived
theoretical and practical bounds on the initial value of skew-compensated clock
based on a systematic analysis of the errors of floating-point operations. The
theoretical bounds provide a reference framework for the analysis and comparison
of bounds on the initial value of skew-compensated clock, while the practical
bounds could replace the approximate bounds suggested in \cite{Kim:22-1} as an
implementation option for limited floating-point precision. The numerical
examples demonstrate that, unlike the approximate bounds of \cite{Kim:22-1}, the
proposed practical bounds based on single-precision floating-point format do not
violate the theoretical bounds and thereby can guarantee the correctness of the
clock skew compensation even on resource-constrained computing platforms like
wireless sensor nodes.

\balance 


\end{document}